\documentclass[%
 reprint,
 amsmath,amssymb,
prd,
]{revtex4-2}

\usepackage{graphicx}
\usepackage{dcolumn}
\usepackage{bm}
\usepackage{amssymb}
\usepackage{subfigure}
\usepackage[utf8]{inputenc}
\usepackage[T1]{fontenc}
\usepackage{booktabs, array, float, tabularx, booktabs, lipsum, amsmath,multirow}
\usepackage{siunitx, xcolor,caption}
\usepackage[version=4]{mhchem}
\usepackage[colorlinks,linkcolor=blue,anchorcolor=blue,citecolor=blue]{hyperref}
\usepackage{multirow}
\newcommand{\hx}[1]{{#1}}

\begin{document}


\title{Anisotropic Lorentz invariance violation in reactor neutrino experiments}


\author{Hai-Xing Lin$^{a}$}
\author{Jie Ren$^{a}$}
\author{Jian Tang$^{a}$}

\affiliation{$^{a}$School of Physics, Sun Yat-Sen University, 510275 Guangzhou, China}

\email{tangjian5@mail.sysu.edu.cn}


\date{\today}
\begin{abstract}
Recent reactor neutrino oscillation experiments reported precision measurements of $\sin^2 2\theta_{13}$ and $\Delta m^2_{ee}$ under the standard 3$\nu$ oscillation framework. However, inter-experiment consistency checks through the parameter goodness-of-fit test \hx{reveal} proximity to tension boundary, with the Double Chooz, RENO, and Daya Bay ensemble yielding $p_\text{PG}=0.14$ vs threshold $\alpha=0.1$. Anisotropic Lorentz invariance violation (LIV) can accommodate this tension by introducing a location-dependent angle $\theta^\text{LIV}$ relative to the earth's axis. It is found that anisotropic LIV improve the fit to data up to 1.9$\sigma$ confidence level significance, with the coefficient $\mathcal{A}^{C(0)}_{31} = 2.43\times 10^{-17}\text{ MeV}$ yielding the best fit, while Parameter Goodness-of-fit (PG) is significantly improved within the LIV formalism.
\end{abstract}


\maketitle


\section{\label{sec:Introduction}Introduction}


It is a great success to describe fundamental particles and their interactions in Standard Model (SM)~\cite{Zyla:2729066}. However, SM does not include gravity, one of the four fundamental interactions, and is incompatible with general relativity. There is a strong quest for new physics beyond SM. Discovery of neutrino oscillations is the first direct experimental evidence beyond SM, pointing to the massive neutrinos and leaving room for imaginations in new physics~\cite{Barger2012}. It is suggested that SM is a low-energy effective theory of a high-energy unified picture at the Planck scale~\cite{Mohapatra1992,Langacker:1980js}. These grand unified theories, such as quantum gravity~\cite{PhysRevD.79.065018,torri2025neutrinospossibleprobesquantum}, string theory~\cite{PhysRevD.39.683,ALANKOSTELECKY1991545} and supersymmetry models~\cite{Berger:2001rm}, have been subject to considerable discussion and often predicted the broken Lorentz symmetry and CPT symmetry, where the product of charge conjugation (C), parity transformation (P) and time reversal (T) are not kept invariant.

CPT symmetry and Lorentz invariance are closely related~\cite{Greenberg:2002uu}. These symmetries are fundamental symmetries of SM in particle physics~\cite{Schwinger:1951xk,LUDERS19571}, as they ensure that physics \hx{stays} the same regardless of the observer. Thus, CPT and Lorentz invariance violation (LIV) suggests the presence of new interactions and gives rise to distinctive phenomena~\cite{sym12111821,Cordero_2025}. {These phenomena manifest not only in astrophysical neutrinos beyond TeV~\cite{Huang_2024,icecubecollaboration2025searchextremelyhighenergyneutrinosconstraints,yang2025constraintslorentzinvarianceviolation}, but also in modifications to neutrino oscillations at the SM scale~\cite{universe6030037}.}

The most popular model in the study of effective field theories for \hx{Lorentz and CPT} violation is the SME (Standard Model Extension), which includes all possible Lorentz invariance violating operators~\cite{Colladay:1996iz,Colladay:1998fq,Kostelecky:2003fs}. Following this model, LIV study has been applied in many neutrino experiments, including MINOS~\cite{PhysRevLett.105.151601,PhysRevLett.101.151601,PhysRevD.85.031101}, LSND~\cite{PhysRevD.72.076004}, {INO}~\cite{Sahoo_2022,Sahoo_2023,Raikwal:2023lzk}, IceCube~\cite{PhysRevD.82.112003},  SuperK~\cite{PhysRevD.91.052003}, {T2K/T2HK}~\cite{PhysRevD.95.111101,Agarwalla_2023,Raikwal:2023lzk,Pan:2023qln}, NO$\nu$A~\cite{Mishra_2024}, MiniBooNE~\cite{AGUILARAREVALO20131303}, Daya Bay~\cite{PhysRevD.98.092013}, Double Chooz~\cite{PhysRevD.86.112009}, {JUNO}~\cite{Li_2014} and   DUNE~\cite{Sarker:2023mlz,mishra2024exploringnonisotropiclorentzinvariance,Agarwalla_2023,Raikwal:2023lzk}. {Among these, reactor neutrino oscillation experiments provide the most precise measurement of standard neutrino oscillation parameters, mainly facilitating tests of the anisotropic LIV mechanism by search for a time-varying neutrino oscillation, while the static anisotropic LIV effects received scant attention.} 

Current reactor neutrino oscillation experiments primarily focus on mixing angle \(\theta_{13}\) and the effect mass squared difference \(\Delta m^2_{ee}\), represented by short-baseline reactor neutrino experiments, such as Double Chooz~\cite{DoubleChooz:2019qbj}, RENO~\cite{RENO:2018dro}, and Daya Bay~\cite{DayaBay:2022orm}. They nailed down the sub-percent precision on \(\theta_{13}\). However, the consistency between the the best-fit results from three reactor neutrino experiments in the standard neutrino oscillation picture demonstrates Parameter Goodness-of-fit~\cite{Maltoni:2003cu} ($p_\text{PG}=0.14$), residing the proximity to the tension boundary. It is a question whether the incompatibility points to new physics beyond three neutrino oscillations. For example, an anisotropy caused by the LIV mechanism could logically lead to a discrepancy, since neutrinos propagate in different directions for these reactor experiments. Then it deserves investigating the effects of anisotropic LIV on the analyses of \(\theta_{13}\) and \(\Delta m^2_{31}\) in the Double Chooz, RENO, and Daya Bay reactor neutrino experiments. This paper aims to further extend the study to anisotropic Lorentz invariance violating phenomena in reactor neutrino oscillation experiments, focusing on inherent directional differences rather than the traditional time-dependent sidereal effects.

The structure of the article is as follows: In Section~\ref{sec:LIV_theory}, we briefly introduce the theoretical framework of anisotropic LIV. In Section~\ref{sec:simulation_method}, we describe the experimental data and numerical methods adopted in this work. Section~\ref{sec:results} presents the impact of anisotropic LIV effects on the analysis of existing short-baseline reactor neutrino oscillation data. In the end, Section~\ref{sec:conclusion} gives the summary and conclusion. 

\section{\label{sec:LIV_theory}Anisotropic Lorentz invariance violation in neutrino oscillation}

We will briefly review the mathematical formulation of neutrino oscillation theory with LIV in this section and provide the effective expression for the survival probability of electron antineutrinos. The theoretical model originates from Ref.~\cite{Kostelecky:2003cr}, while the simplification of effective parameter is similar to Ref.~\cite{Kostelecky:2004hg}. Since the mass of neutrinos is extremely small, smaller than 0.8 eV based on the latest KATRIN results~\cite{KATRIN:2021uub}, the LIV energy-momentum dispersion relation can be expressed under extreme relativistic approximation \cite{PhysRevD.59.116008}: 
\begin{equation}\label{eq:Energy-moment_disperation}
    E \simeq |\bm{p}|+ \frac{m^2}{2|\bm{p}|} + \sum_{n=1}^{\infty}\gamma_{\text{LIV},n}|\bm{p}|^{n-1}.
\end{equation}
In Eq.~(\ref{eq:Energy-moment_disperation}), the second term on the right-handed side is a perturbative expansion of the neutrino mass term, and the third term predicts subtle \hx{LIV} suppressed by the Planck scale $E_\text{Pl}\simeq10^{19}\text{ GeV}$~\cite{Addazi_2022}. Ignoring neutrino-antineutrino mixing, the LIV Lagrangian in the neutrino sector of SME is given by~\cite{PhysRevD.70.031902,Kostelecky:2011gq}:
\begin{equation}\label{eq:LIV_Lagrangian}
\begin{aligned}
    \mathcal{L}_{D}=&-\overline{\nu}_\alpha(a^{\mu}\gamma_\mu+b^\mu\gamma_5\gamma_\mu)\nu_\beta\\
    +&\hx{\frac{i}{2}\overline{\nu}_\alpha(c^{\mu\nu}+d^{\mu\nu}\gamma_5)\gamma_\mu\overset{\leftrightarrow}{\partial}_\nu\nu_\beta}.
\end{aligned}
\end{equation}
The \hx{subscripts} \( \alpha,\beta = e, \mu, \tau \) of the neutrino field correspond to different lepton flavors. If we assume that the mass eigenstates of neutrinos are the physical states propagating with LIV in vacuum, the effective Hamiltonian for neutrino oscillations in flavor eigenstates with LIV can be expressed as \cite{PhysRevD.77.013007}:
\begin{equation}\label{eq:LIVosc_Hamiltonian_mass}
    \mathcal{H}_{\nu\nu}^\text{eff} = U^\dagger(\frac{M^2}{2E}+\mathcal{H}_\text{LIV})U + V_\text{matter}.
\end{equation}
The unitary matrix \( U \) is the PMNS matrix \cite{Pontecorvo:1967fh}, and \( V_\text{matter} \) is the matter potential term. After removing the \hx{LIV} term, \((\mathcal{H}_{\nu\nu})_\text{eff}\) is the effective Hamiltonian for standard neutrino oscillations. In the leading-order perturbative expansion approximation, considering the Sun-centered reference frame with the Earth's rotation axis as the Z-axis~\cite{Kostelecky:2004hg}, the Hamiltonian term $\mathcal{H}_\text{LIV}^\text{eff} =\textbf{Diagonal}(0,\Delta\gamma_{21},\Delta\gamma_{31})$ causing anisotropic LIV can be reconstructed into a form depending on directional factors,
\begin{widetext}
\begin{equation}\label{eq:LIV_Hamiltonian_eff}
\begin{aligned}
    \Delta \gamma_{ij} = (\mathcal{A}^{S(0)}_{ij}+\mathcal{A}^{S(1)}_{ij}E)\sin\theta^\text{LIV}+(\mathcal{A}^{C(0)}_{ij}+\mathcal{A}^{C(1)}_{ij}E)\cos\theta^\text{LIV}+\mathcal{B}^{S}_{ij}E\sin2\theta^\text{LIV}+\mathcal{B}^{C}_{ij}E\cos2\theta^\text{LIV},
\end{aligned}
\end{equation}
\end{widetext}
where the $\mathcal{A},\mathcal{B}$ parameters are effective LIV coefficients in general form for a several year experiment. $\mathcal{A}^{(0)}$ would undergo a sign inversion by CPT transformation in the antineutrino case. The LIV correction term depends only on the polar angle \(\theta^\text{LIV}\) in the direction. \hx{The relation between SME coefficients and effective anisotropic LIV parameters can be found in the appendix~\ref{appendix:A}.}

\begin{figure}[htbp]
    \centering
    \includegraphics[width = 0.5\textwidth]{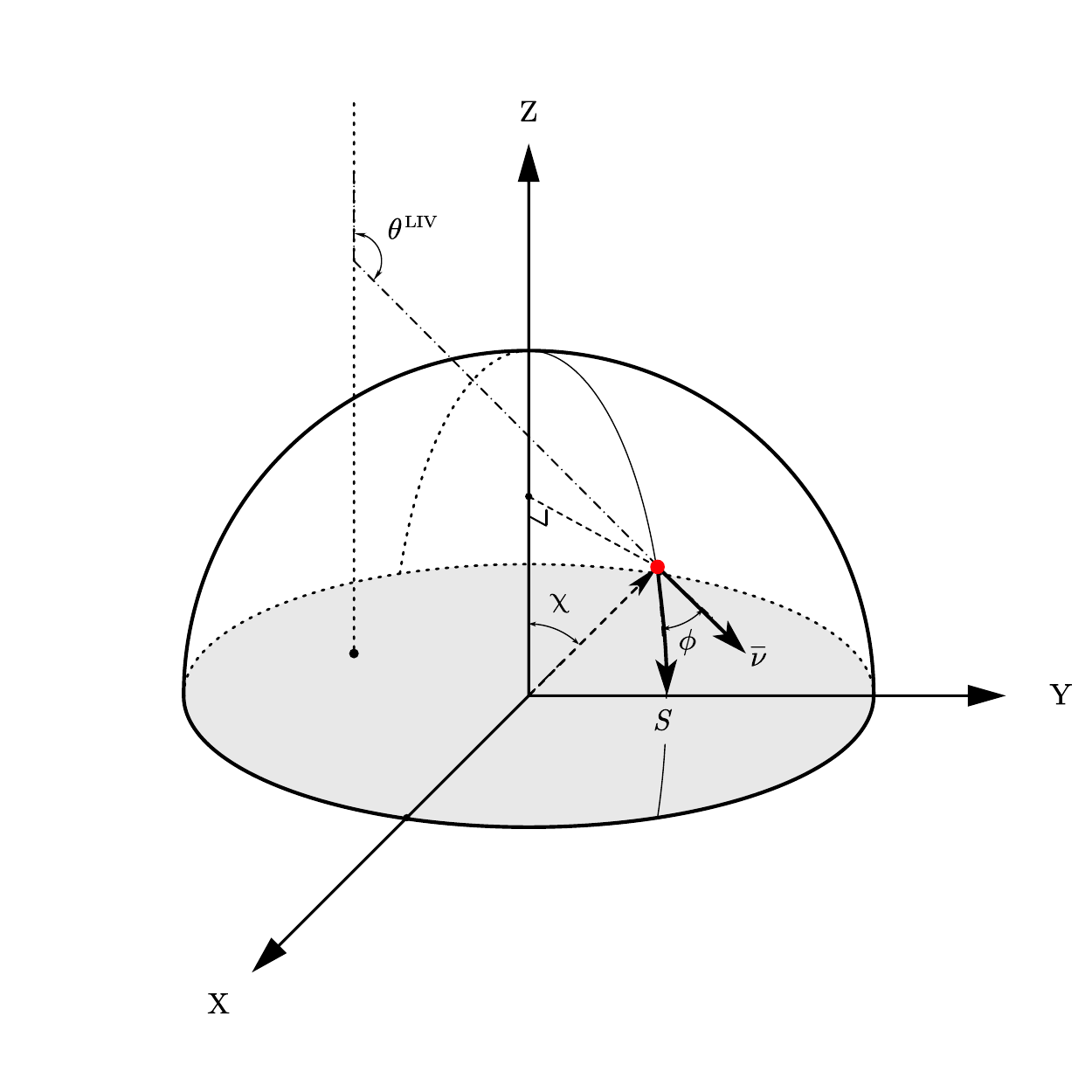}
    \caption{The sketch to show how antineutrinos propagate along the Earth's surface. The silhouette of Earth's northern hemisphere is with the rotation axis as the Z-axis. The red dot represents the geographical location of the detector, while the label ``S'' marks the due south at this point.}
    \label{fig:LIV_angle_south}
\end{figure}

$\theta^\text{LIV}$ is the angle betwenn the neutrino beam and the Earth's rotation axis, given by
\begin{equation}\label{eq:theta_LIV}
    \theta^\text{LIV}=\arccos(-\sin\chi\cos\phi),
\end{equation}
as zenith angle approximates right angle for reactor neutrino experiments. $\chi$ is the co-latitude of the location of the detector, while $\phi$ is the angle between the neutrino beam and the south measured towards the east. $\cos\phi$ reflects the projection of the beam in the due south direction, which contributes to the axial component, as shown in Fig.~\ref{fig:LIV_angle_south}. Since eastward or westward displacement does not alter this axial projection, \(\theta^\text{LIV}\) does not change by the Earth's rotation without inducing a time-dependent sidereal effect. 

The survival probability of reactor electron antineutrinos with anisotropic LIV in short-baseline reactor experiments can be written as follows,
\begin{equation}
\begin{aligned}
    P_{\bar{\nu}_e\rightarrow\bar{\nu}_e} \approx& 1 - \sin^2 2\theta_{13} \sin^2 [(\frac{\Delta m^2_{ee}}{4E}+\frac{\Delta \gamma_{31}}{2})L] \\
    &- \cos^4 \theta_{13} \sin^2 2\theta_{12} \sin^2 [(\frac{\Delta m^2_{21}}{4E}+\frac{\Delta \gamma_{21}}{2})L] .
\end{aligned}
\end{equation}
Here, $\Delta m_{ee}^2=\cos^2 \theta_{12} \Delta m^2_{31} + \sin^2 \theta_{12} \Delta m^2_{32}$.

\hx{In the LIV framework, neutrino oscillations acquire linear phase terms with energy dependence, which grow in magnitude with increasing energy in contrast to standard neutrino oscillation phases $\Delta m^2 L/E$, thereby allowing for a more flexible fit to the reconstructed neutrino energy spectrum. Moreover, $\theta^\text{LIV}$ in $\Delta \gamma$ depends on specific propagation direction, leading to LIV phase variations across different experiments and resulting in inconsistent survival probabilities even at identical energy and baseline. Thus, these subtle LIV-induced differences may help explain the discrepancies between the best-fit results of short-baseline reactor neutrino experiments when LIV effects are neglected.}

\section{Simulation Details\label{sec:simulation_method}}

In the present work, we focus on data analysis from short-baseline reactor experiments Double Chooz, RENO and Daya Bay. These experiments relied on large flux of $\bar{\nu}_e$ from reactors and detected $\bar{\nu}_e$ in liquid scintillator (LS) by inverse-beta-decay (IBD). Then the energy of $\bar{\nu}_e$ is inferred from a prompt-energy ($E_{\bar{\nu}} \sim E_\text{prompt}+0.78\text{ MeV}$). Recent published data from Double Chooz, RENO, and Daya Bay \cite{Soldin:2024fgt,Shin:2020mue,DayaBay:2022orm} is reproduced within the GLoBES framework \cite{Huber:2004ka,Huber:2007ji,DayaBay:2015lja,Berryman_2021}. Main configuration for each experiment is shown in Table~\ref{tab:exp_configuration} \footnote{The geographic information for each experiment was partially referenced from Google Maps results.}.

Since the results from GLoBES are consistent with that from reference, we \hx{investigate} the effects of anisotropic LIV on the measurements of \(\theta_{13}\) and \(\Delta m^2_{ee}\). In this section, we briefly introduce the mock data preparations in each experiment and present the data analysis methods used in the study. Detailed computational procedures and their implementations can be accessed through the public GitHub repository \cite{Lin_ReactorLIV_2023}.

\begin{table*}[htbp!]
	\centering
	\begin{tabular}{cccccccc}
	\hline
        \hline
\multirow{2}{*}{Experiment} & \multicolumn{2}{c}{DoubleChooz} &  \multicolumn{2}{c}{RENO} & \multicolumn{3}{c}{DayaBay}  \\
\cmidrule{2-8}
& ND & FD & ND & FD & EH1 & EH2 & EH3 \\
\midrule
Source & \multicolumn{2}{c}{France} &  \multicolumn{2}{c}{Korea} &  \multicolumn{3}{c}{China}  \\
Power [$\mathrm{GW}_\mathrm{th}$]   & \multicolumn{2}{c}{$4.25\times 2$} &  \multicolumn{2}{c}{$2.8\times 6$} &  \multicolumn{3}{c}{$2.9\times 6$} \\     
LS mass & $10.6\,$ t & $10.6\,$ t & $15.4\,$ t & $15.4\,$ t & $40.0\,$ t & $40.0\,$ t & $80.0\,$ t  \\
Baseline &$400\,$ m &$1050\,$ m &$300\,$ m & $1400\,$ m & $512\,$ m & $561\,$ m & $1579\,$ m   \\
Data source & \multicolumn{2}{c}{Ref.~\cite{Soldin:2024fgt}} & \multicolumn{2}{c}{Ref.~\cite{Shin:2020mue,RENO2022}} & \multicolumn{3}{c}{Ref.~\cite{DayaBay:2022orm}}  \\
$\theta^\mathrm{LIV}$ & \multicolumn{2}{c}{$1.86$} & \multicolumn{2}{c}{$2.25$} & \multicolumn{3}{c}{$0.74$}  \\
    \hline
    \hline
\end{tabular}
\caption{Summary of reactor neutrino experimental configurations\label{tab:exp_configuration}}
\end{table*}

\subsection{Double Chooz experiment\label{DoubleChooz}}
The Double Chooz experiment made good use of Chooz reactors, which are located at Chooz in France with a co-latitude of $\chi = 39.9^\circ$, with total power of 8.5 GW\(_{th}\). The near detector (ND) and the far detector (FD) are located approximately at 400 m and 1050 m from reactor cores, respectively. {The} far detector was put to the east of the reactors at an angle of $\phi=63.4^\circ$, corresponding to $\theta^\text{LIV} = 1.86$ rad.

The core of the Double Chooz detector contained $10.3 \, ~\text{m}^3$ of LS with $0.1\%$ Gd concentration \cite{DoubleChooz:2022ukr}. According to the latest Double Chooz data \cite{Soldin:2024fgt}, the near detector collected 412k IBD signals over 587 days, while the far detector took data of 125k IBD signals over 1276 days. We intend to utilize all data in analysis to examine new physics. The energy binning for ND and FD is identical, with the positron events divided into 38 bins in the energy range $[1.0, 20.0] \, \text{MeV}$. 

\subsection{RENO experiment\label{sec:RENO}}
The full name of the RENO experiment is the Reactor Experiment for Neutrino Osscillation, located near the Hanbit Nuclear Power Plant in Korea yielding a total thermal power of $6\times 2.8 \, \text{GW}_{\text{th}}$. RENO employed a dual detector design similar to that of Double Chooz, with two detectors located approximately 300 m and 1400 m from the reactors, respectively. The far detector of RENO was placed at $\phi = 39.2^\circ$ to the southeast of the reactors, with a co-latitude of $\chi = 54.6^\circ$, resulting in $\theta^\text{LIV} = 2.25$ rad.

The neutrino target of the RENO detector is similar in material to that of the Double Chooz, consisting of 16.5 tons of hydrocarbon LS mixed with $0.1\%$ Gd concentration~\cite{RENO:2016ujo}. Our simulation analyzes about 120k IBD signal data from the far detector over approximately 2900 days until 2022, and includes around 990k data from the near detector over about 2500 days~\cite{Shin:2020mue,RENO2022}, which are divided into 57 bins in the range $[1.2, 8.4] \, ~\text{MeV}$. We investigate neutrino oscillation by the data from the far detector divided into 45 bins in the energy range $[1.2, 8.0] \, ~\text{MeV}$.

\subsection{Daya Bay experiment\label{sec:DayaBay}}
Daya Bay experiment is near the Daya Bay Nuclear Power Plant. There are three pairs of reactors, each with a power of 2.9 GW$_{\text{th}}$. The reactor neutrino experiments took a further step by placing detectors in three different orientations. The three experimental halls EH1, EH2, and EH3, were arranged with a baseline of 512 m, 561 m, and 1579 m from the reactor cores, respectively~\cite{DayaBay:2013yxg}. The farthest EH3 was located at the northwest of the reactor with $\phi = 216.6^\circ $ at $\chi = 67.4^\circ$, from which the LIV-related angle is calculated to be $\theta^\text{LIV}_{\text{DayaBay}} = 0.74$ rad.

Analogous to the Double Chooz and RENO detectors, the core of the Daya Bay detectors consisted of 20 tons of LS with $0.1\%$ Gd concentration. We reproduced the data in the simulation tool with approximately 2.2 million IBD rates allocated to EH1, 2.5 million to EH2, and 764k to EH3, respectively. The data samples are divided into 26 bins in the energy range $[0.7, 12.0] \, \text{MeV}$. In the analysis, all data from the three experimental halls are used to quantify the neutrino oscillation and its new physics potentials.

\subsection{Statistical method\label{subsec:numerical_method}}
The data analysis is based on the likelihood method for each experiment with the $\chi^2$ function defined as follows,
\begin{equation}\label{eq:LIV-pull-method}
\begin{aligned}
        &\chi^2_\text{exp}(\theta_{13},\Delta m^2_{ee},\Delta \gamma) \\
        &= \text{min}_{\{\xi\}}\left[ \sum_i \left(\frac{(O_{i,d}-T_{i,d}(\{\xi\}))^2}{O_{i,d}}\right) + \sum_k \frac{\xi_k^2}{\sigma_k^2}\right]\,,
\end{aligned}
\end{equation}
where \( O_{i,d} \) is the observed events for each detector in each experiment, while the subscript \( i \) iterates over all energy bins, as \( d \) represents different detectors. 
Theoretically predicted events \( T_{i,d} \) are represented by  $T^\text{nosc}_{i,d}$ without oscillation,  based on experimental configurations given in Table.~\ref{tab:exp_configuration}. 
\hx{Our simulations were carried out using GLoBES~\cite{Huber:2004ka,Huber:2007ji}, with anisotropic LIV effects implemented in its neutrino oscillation calculation engine, where $\sin^2\theta_{12}=0.304$ and $\Delta m^2_{21}=\qty{7.42e-5}{eV^2}$~\cite{Esteban_2020} are used as input to the fit. Additionally, due to the short baseline, the uncertainties of matter density and $\theta_{12},\Delta m^2_{21}$ can be neglected.}


Meanwhile, the systematic uncertainties are introduced with a set of nuisance parameters ${\xi}$ by the pull method \cite{PhysRevD.66.053010}, \hx{where normalization and energy calibration errors are taken into account. The normalization errors include reactor neutrino flux, detection efficiency and energy scale uncertainties, and the energy calibration error refers to the precision of neutrino energy reconstruction. These systematic errors are implemented via AEDL rule definitions in GLoBES~\cite{Huber:2020GLoBESManual3.2.18}, while
the values of the systematical uncertainties \( \sigma_k \) are \hx{adopted} from Ref.~\cite{DoubleChooz:2019qbj,RENO:2018dro,DayaBay:2022orm}.} 

In total, 12 LIV parameters are tested sequentially in terms of both Standard Gooness-of-fit (SG) and Parameter Goodness-of-fit (PG)~\cite{Maltoni:2003cu,Esteban_2020}. \hx{SG} is evaluated by the total $\chi^2$,
\begin{equation}
    \chi^2_\text{SG} = \chi^2_\text{Double Chooz}+\chi^2_\text{RENO}+\chi^2_\text{Daya Bay},
\end{equation}
$\chi^2_\text{SG,SM}$ \hx{corresponding} to standard neutrino oscillations is a special case with the LIV parameter $\Delta\gamma=0$. Thus, we can test an extended LIV term by 
\begin{equation}\label{eq:SG_LRT}
    \Delta\chi^2_\text{SG} = \chi^2_\text{SG,SM} - \chi^2_\text{SG},
\end{equation}
asymptotically following a chi-square function with one degree of freedom. It is noteworthy that we have fixed $\Delta m^2_{ee} = 2.519\times10^{-3}\text{ eV}^2$ in \hx{$\chi^2$ of the Double Chooz, since it only provided a precise measurement of $\sin^22\theta_{13}$} in the reference~\cite{DoubleChooz:2019qbj}. On the other hand, we have \hx{PG} by
\begin{equation}
    \chi^2_\text{PG}(\theta_{13},\Delta m^2_{ee},\Delta \gamma_\text{LIV}) = \chi^2_\text{SG}-\sum_r^\text{exp}\chi^2_{r,\text{min}}.
\end{equation}
$\chi^2_{r,\text{min}}$ is the minimum with the data only from a single experiment, while $\chi^2_\text{SG}$ is involved with all three experiments. The p-value associated with the $\chi^2_\text{PG}$
in statistics can be calculated for both the standard neutrino oscillation and the LIV scenario, taking into account their respective degrees of freedom (3 and 5).

By computing $\Delta \chi^2_\text{SG}$, we evaluate the extent to which introducing a LIV parameter improves the global fit of the theoretical prediction confronted with the combined experimental results. However, given that the three reactor experiments exhibit distinct sensitivities to LIV due to their different configurations represented by $\theta^\text{LIV}$, the statistic measure $\chi^2_\text{PG}$ becomes essential to rigorously assess the inter-experiment consistency of the LIV parameters. Our analysis aims to demonstrate that the LIV-modified oscillation framework not only provides a superior phenomenological description of data but also fulfills the universality requirement across reactor experiments, thereby supporting its validity as a generalized extension in new physics searches.

\section{Numerical results\label{sec:results}}
\begin{table*}[htbp]
\centering
\begin{tabular}{c c c c c | c c}
\hline
\hline
Parameter & LIV value & $\sin^22\theta_{13}$ & $\Delta m^2_{ee},10^{-3} \text{eV}^2$ & $\Delta\chi^2_\text{SG}$ & $\chi^2_\text{PG}$  & p-value \\
\hline
SM & - & 0.0873 & 2.535 & - & 5.55 & 0.14 \\
$\mathcal{A}^{S(0)}_{21}$ & 1.17$\times10^{-17}\text{ MeV}$ & 0.0889 & 2.503 & 0.23 & 6.37 & 0.27 \\
$\mathcal{A}^{S(1)}_{21}$ & -3.34$\times10^{-18}\text{}$ & 0.0889 & 2.511 & 0.77 & 5.95 & 0.31 \\
$\mathcal{A}^{C(0)}_{21}$ & -7.44$\times10^{-20}\text{ MeV}$ & 0.0873 & 2.536 & $\simeq0$ & 6.61 & 0.25 \\
$\mathcal{A}^{C(1)}_{21}$ & -2.69$\times10^{-18}\text{}$ & 0.0880 & 2.518 & 0.62 & 6.29 & 0.28 \\
$\mathcal{B}^{S}_{21}$ & -2.21$\times10^{-18}\text{}$ & 0.0880 & 2.516 & 0.78 & 6.13 & 0.29 \\
$\mathcal{B}^{C}_{21}$ & -2.92$\times10^{-17}\text{}$ & 0.0874 & 2.533 & 0.17 & 6.74 & 0.24 \\
$\mathcal{A}^{S(0)}_{31}$ & -1.88$\times10^{-17}\text{ MeV}$ & 0.0877 & 2.459 & 1.00 & 6.53 & 0.26 \\
$\mathcal{A}^{S(1)}_{31}$ & 7.70$\times10^{-19}\text{}$ & 0.0875 & 2.525 & 0.10 & 6.77 & 0.24 \\
$\mathcal{A}^{C(0)}_{31}$ & 2.43$\times10^{-17}\text{ MeV}$ & 0.0866 & 2.621 & 3.67 & 3.87 & 0.57 \\
$\mathcal{A}^{C(1)}_{31}$ & -3.44$\times10^{-18}\text{}$ & 0.0866 & 2.572 & 2.06 & 4.82 & 0.44 \\
$\mathcal{B}^{S}_{31}$ & -2.50$\times10^{-17}\text{}$ & 0.0866 & 2.570 & 2.16 & 4.72 & 0.45 \\
$\mathcal{B}^{C}_{31}$ & -3.00$\times10^{-17}\text{}$ & 0.0873 & 2.538 & 0.20 & 6.68 & 0.25 \\
\hline
\hline
\end{tabular}
\caption{Fit results by minimizing $\chi^2_\text{SG}$ and $\chi^2_\text{PG}$ with each anisotropic LIV parameter. $\chi^2_\text{PG}$ is characterized by 3 d.o.f. in standard neutrino oscillation and 5 d.o.f. under the LIV scenario. \label{tab:result_SG_PG}}
\end{table*}
In this work, we investigate the effect of the anisotropic LIV in the leading-order perturbation without a loss of generality. Numerical results based on $\Delta\chi^2_\text{SG}$ and $\chi^2_\text{PG}$ are presented in light of 12 anisotropic LIV parameters, which can be sequentially examined. It is essential to check whether LIV can accommodate the tension between different experiments and improve the global fit to data in three reactor neutrino oscillation experiments.
\begin{figure}[htbp]
    \centering
    \includegraphics[width=0.5\textwidth]{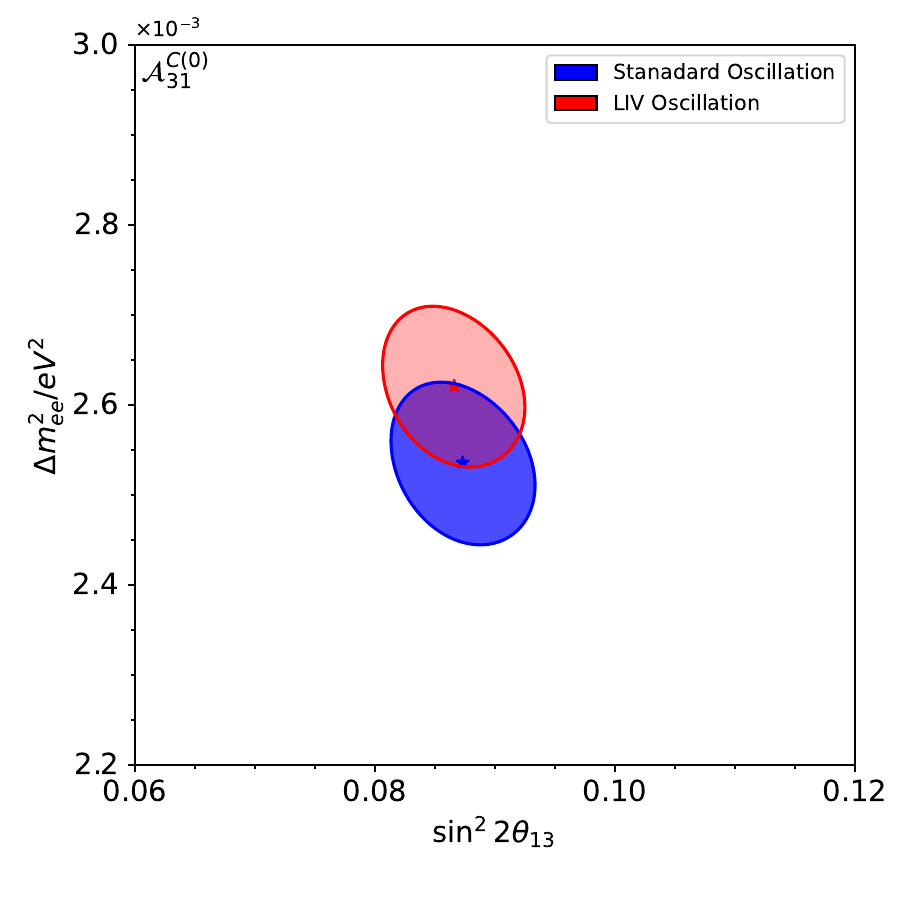}
    \caption{Shifts of neutrino oscillation parameters under LIV effects. The red and blue correspond to the \hx{90\% C.L.} analysis results of standard neutrino oscillation and LIV oscillation respectively, assuming normal mass hierarchy.}
    \label{fig:LIV_2D_combined}
\end{figure}
SG are PG complementary to examine the impact of LIV parameters, while the best-fit \hx{points} in $\chi^2_\text{SG}$ and $\chi^2_\text{PG}$ are identical. The fit results of $\sin^22\theta_{13}$, $\Delta m^2_{ee}$ and the corresponding LIV values are presented in Table~\ref{tab:result_SG_PG}.

\begin{figure}[htbp]
    \centering
    \includegraphics[width=0.5\textwidth]{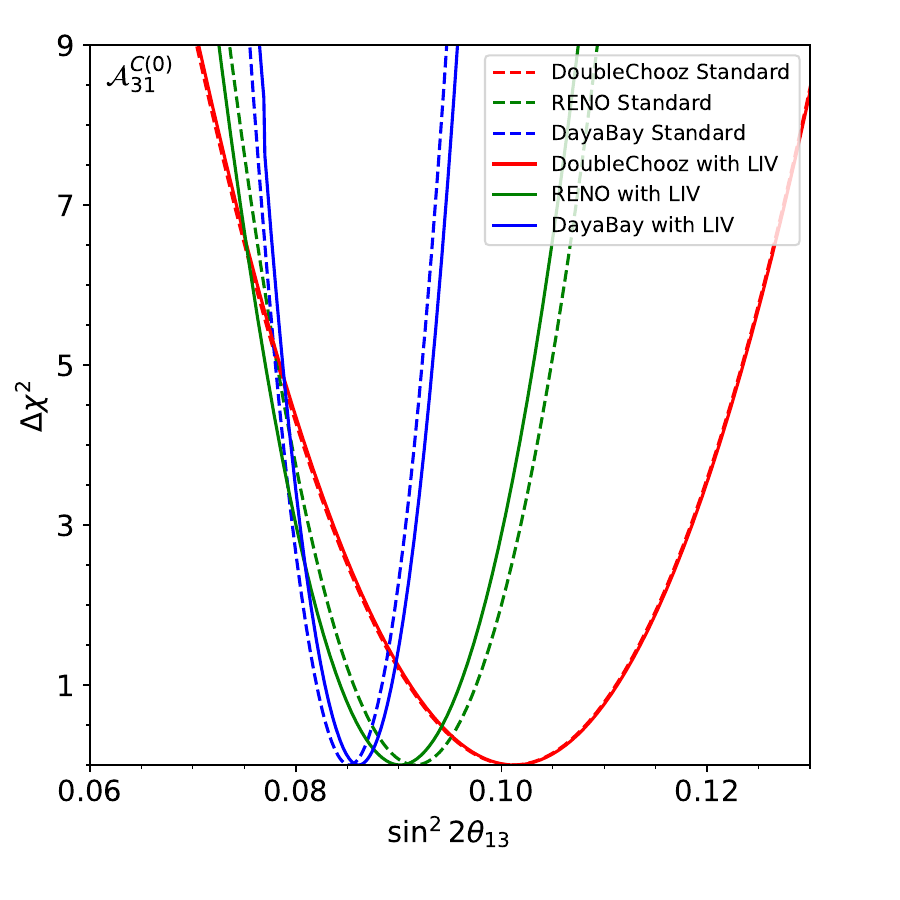}
    \caption{$\chi^2$ curves for $\sin^2 2\theta_{13}$ analysis of each experiment after introducing the LIV parameter $\mathcal{A}^{C(0)}_{31}=2.43\times10^{-17}\text{ MeV}$. The red, green, and blue colors correspond to the Double Chooz, RENO, and Daya Bay experiments, respectively. The solid lines represent the $\chi^2$ curves under the effect of LIV, while the dashed lines are in the standard neutrino oscillation case. $\Delta m^2_{ee}$ is fixed at $2.519\times10^{-3}\text{ eV}^2$, assuming normal mass hierarchy.\label{fig:LIV_th13_each}}
\end{figure}

For $\chi^2_\text{SG}$, LIV parameters $\mathcal{A}^{S(0)}_{31}$, $\mathcal{A}^{C(0)}_{31}$, $\mathcal{A}^{C(1)}_{31}$ and $\mathcal{B}^{S}_{31}$ exhibit an enhancement beyond $1\sigma$, with $\mathcal{A}^{C(0)}_{31}$, $\mathcal{A}^{C(1)}_{31}$ and $\mathcal{B}^{S}_{31}$ to demonstrate more pronounced effects, while $\mathcal{A}^{C(0)}_{31}$ yields the largest enhancement of about $1.9\sigma$ confidence level (C.L.) in standard goodness of fit. The anisotropic LIV parameters cause very little shift in the fitting results of $\sin^22\theta_{13}$ and $\Delta m^2_{ee}$, where their impact is represented by $\mathcal{A}^{C(0)}_{31}=2.43\times10^{-17}\text{ MeV}$ shown in Fig~\ref{fig:LIV_2D_combined}.

\begin{figure}[htbp]
    \centering
    \includegraphics[width=0.5\textwidth]{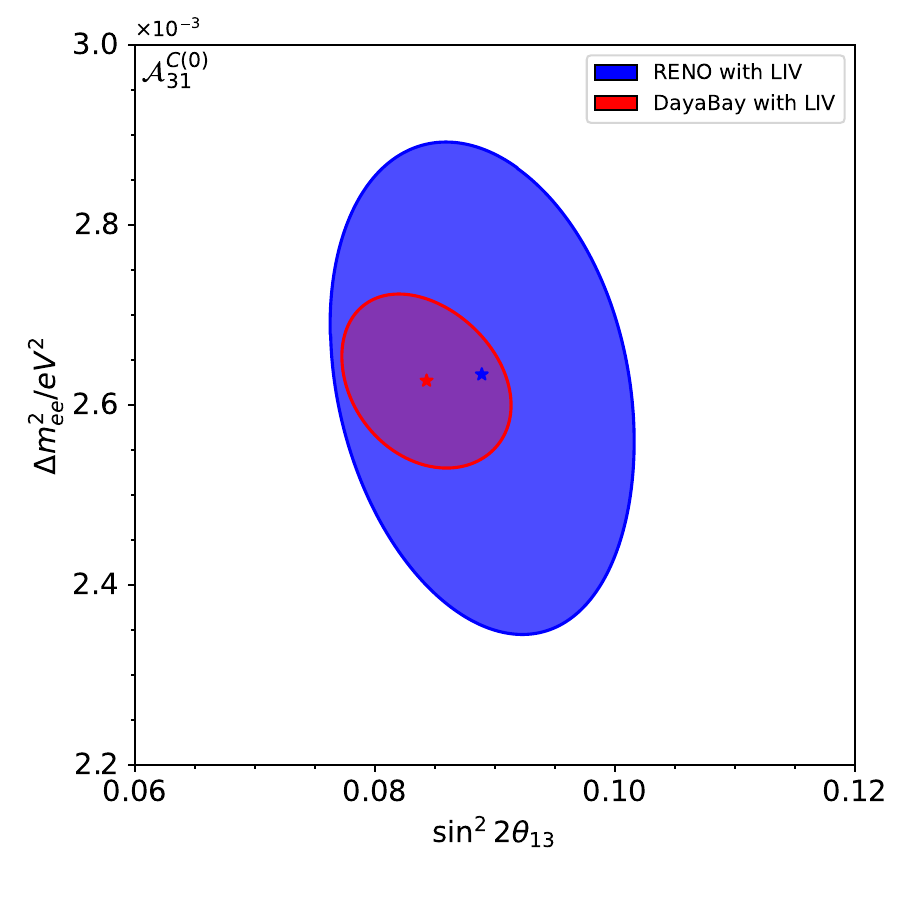}
    \caption{90\% C.L. space of neutrino oscillation parameters for the RENO and Daya Bay experiments after introducing $\mathcal{A}^{C(0)}_{31}=2.43\times10^{-17}\text{ MeV}$. The red and blue colors represent the results of Daya Bay and RENO, respectively. The star markers indicate the best-fit points, assuming normal neutrino mass hierarchy. \hx{Double Chooz is excluded owing to its absence of a precise $\Delta m^2_{ee}$ measurement~\cite{DoubleChooz:2019qbj}.}}
    \label{fig:LIV_2D_each}
\end{figure}

When it comes to $\chi^2_\text{PG}$, we have p-value to quantify \hx{PG} before and after an introduction of each LIV parameter. In the standard neutrino oscillation framework, $\chi^2_\text{PG}$ is characterized by three degrees of freedom, whereas in LIV scenario, it extends to five degrees of freedom. Bearing an increase of degrees of freedom in mind, we focus on the LIV parameters $\mathcal{A}^{C(0)}_{31}$, $\mathcal{A}^{C(1)}_{31}$ and $\mathcal{B}^{S}_{31}$ that demonstrate the most significant improvements. When adopting $\alpha = 0.1$ as the compatibility threshold, the standard neutrino oscillation with $p=0.14$ barely satisfies the threshold requirement, while the LIV framework with $\mathcal{A}^{C(0)}_{31}$ achieves high confidence level at $p=0.57$. The consistency across experimental data samples from short-baseline reactor neutrino experiments is improved significantly, as shown in Fig~\ref{fig:LIV_th13_each} and \ref{fig:LIV_2D_each}. The fit results of RENO and Daya Bay tend to unify, while the result of Double Chooz is almost the same as the result given by the standard neutrino oscillation. \hx{Due to the coupling between $\mathcal{A}_C$ and $\cos \theta^\text{LIV}$, Double Chooz remains relatively insensitive to LIV effects since its $\theta^\text{LIV}$ is close to $\pi/2$. In contrast, RENO and Daya Bay each exhibit appreciable $\cos \theta^\text{LIV}$ value with opposite signs, leading to antithetical LIV effects that help reconcile the discrepancies among the results.}

The best-fit LIV parameters in $\Delta\gamma$ are almost at the same order consistent with the previous experimental limits~\cite{Kostelecky:2008ts}, including those published by the reactor neutrino experiments Daya Bay \cite{PhysRevD.98.092013} and Double Chooz \cite{PhysRevD.86.112009}.

\section{Conclusion\label{sec:conclusion}}

Lorentz invariance violation has profound impacts on the data analysis and physics interpretation in the laboratory frame, which may not only lead to periodic variations in the reactor neutrino survival probability but also induce discrepancies between the results of different experiments. In the present work, the latest data from short-baseline reactor neutrino experiments were used to explore the effects of anisotropic LIV. We analysed dataset from Double Chooz, RENO, and Daya Bay by incorporating the LIV mechanism into the mass eigenstates of neutrinos. 

Our results indicated that anisotropic LIV model provides a better fit to the total data with respect to standard neutrino oscillation up to 1.9$\sigma$ C.L., with $\mathcal{A}^{C(0)}_{31} = 2.43\times 10^{-17}\text{ MeV}$, $\sin^22\theta_{13} = 0.0866$, $\Delta m^2_{ee} = 2.621\times10^{-3}\text{ eV}^2$ yielding the best-fit. Meanwhile, the LIV $\mathcal{A}^{C(0)}_{31}$ demonstrated the best parameter goodness-of-fit, resolving the tension between short-baseline reactor neutrino experiments data ($p_\text{PG}=0.57$), whereas the standard neutrino oscillation fit remains proximity to the tension boundary $(p_\text{PG}=0.14)$ with an acceptance criterion of $\alpha=0.1$. The inclusion of LIV effects caused minimal changes to the best-fit values of \(\sin^22\theta_{13}\) and \(\Delta m^2_{ee}\) for the overall dataset, while the results from RENO and Daya Bay show convergence.

We anticipate that the next-generation neutrino oscillation experiments like Jiangmen Underground Neutrino Observatory (JUNO)~\cite{JUNO:2021vlw} will be able to examine such LIV effects soon.
\section{Note Added}
During the editing process, the updated results from the RENO experiment \cite{RENO:2024msr} indicate observable deviations from the dataset utilized in this analysis. While the conclusions derived from {the data released by the RENO experiment until 2020~\cite{Shin:2020mue}} remain valid, the newly reported measurements suggest a mitigated sensitivity to the anisotropic LIV effect.
\section{Acknowledgments}
This project was supported in part by National Natural Science Foundation of China under Grant Nos. 12347105. This work was supported in part by Fundamental Research Funds for the Central Universities (23xkjc017) in Sun Yat-sen University. JT is grateful to Southern Center for Nuclear-Science Theory (SCNT) at Institute of Modern Physics in Chinese Academy of Sciences for hospitality. We appreciate Sampsa Vihonen's early-stage involvement in the project and extend gratitude to Yinyuan Huang for his coding contributions.
 
\appendix
\section{The relation between SME coefficients and effective anisotropic LIV \hx{parameters}}
\label{appendix:A}
In the leading-order perturbative expansion approximation, the effective Hamiltonian with SME coefficents can be written as,
\begin{equation}\label{eq:LIV_Hamiltoninian_massindication}
    (\mathcal{H}_\text{LIV})_{ij} = \frac{\delta_{ij}}{E}(a_L^{\mu}p_{\mu}-c_L^{\mu\nu}p_{\mu}p_{\nu})_{ij},
\end{equation}
\( i, j = 1, 2, 3 \) correspond to different mass states of neutrinos. The LIV coefficients in mass eigenstates have a linear correspondence with the coefficients in Equation~(\ref{eq:LIV_Lagrangian}),
\begin{equation}\label{eq:LIV_Hamiltonian}
\begin{aligned}
    (a^\mu_L)_{ij} &= U_{i\alpha}(a^\mu+b^{\mu})_{\alpha\beta}(U^\dagger)_{\beta j},\\
    (c^{\mu\nu}_L)_{ij} &= U_{i\alpha}(c^{\mu\nu}+d^{\mu\nu})_{\alpha\beta}(U^\dagger)_{\beta j}
\end{aligned}
\end{equation}
We assume that the mass states of neutrinos during propagation are physical states, without mixing effects between mass states. Diagonal elements of the Hamiltonian are real numbers required by hermiticity. Therefore, \(a_{ij}\) and \(c_{ij}\) can be regarded as \(3 \times 3\) diagonal real matrices, with mass dimensions of 1 and 0, respectively. For antineutrinos, \(a_L \rightarrow -a_L\).

As mentioned in Eq.~(\ref{eq:LIV_Hamiltonian_eff}), the Hamiltonian term causing anisotropic LIV can be reconstructed into a form dependent on directional factors,
\begin{widetext}
\begin{equation}
    \begin{aligned}
        (\mathcal{H}_\text{LIV})_{ij} =& \left[a^X_L \cos\omega t +
  a^Y_L \sin \omega t - 2(c^{TX}_L  \cos \omega t  + c^{TY}_L  \sin \omega t)E \right]\sin\theta^{\text{LIV}}\\
        &+ \left[a^Z_L - 2c^{TZ}_LE\right]\cos\theta^{\text{LIV}}\\
        &-(c^{XZ}_L\cos\omega t+c^{YZ}_L\sin\omega t)E\sin2\theta^{\text{LIV}}\\
        &+\frac{1}{2}\left[(-c^{ZZ}_L+\frac{1}{2}c^{XX}_L+\frac{1}{2}c^{YY}_L)+\frac{1}{2}(c^{XX}_L-c^{YY}_L)\cos2\omega t+c^{XY}_L\sin2\omega t\right]E\cos2\theta^{\text{LIV}}.
    \end{aligned}
\end{equation}
\end{widetext}
Here, \(\theta^{\text{LIV}}\) is the same as that in Equation~(\ref{eq:LIV_Hamiltonian_eff}), and \(\phi^\text{LIV} = \omega t\) corresponds to the azimuthal angle in the XY plane of this Sun-center reference frame, while the zenith angle is approximated as $\pi/2$. Following the form of mass-squared differences term in standard neutrino oscillation Hamiltonian, the effective LIV Hamiltonian can be expressed as,
\begin{equation}\label{eq:LIV_Hamiltonian_eff2}
\begin{aligned}
\mathcal{H}_\text{LIV}^\text{eff} &=\mathrm{\textbf{Diagonal}}(0,\Delta\gamma_{21},\Delta\gamma_{31}),\\
    \Delta \gamma_{ij} &=\frac{1}{E}\left[(a_L^{\mu}p_{\mu}-c_L^{\mu\nu}p_{\mu}p_{\nu})_{ii}-(a_L^{\mu}p_{\mu}-c_L^{\mu\nu}p_{\mu}p_{\nu})_{jj}\right].
\end{aligned}
\end{equation}
This expression is equivalent to Eq.~(\ref{eq:LIV_Hamiltoninian_massindication}) for neutrino oscillations. In neutrino experiments conducted over several years, the effects of the azimuthal angle \(\phi^{LIV}\) can be represented by the mean value of the periodic function, retaining only the directional dependence on the polar angle $\theta^{LIV}$. After integration, the general form of the anisotropic LIV Hamiltonian in Eq.~(\ref{eq:LIV_Hamiltonian_eff}) can be obtained,
\begin{equation*}
    \Delta \gamma(a^\mu,c^{\mu\nu})\rightarrow\Delta \gamma(\mathcal{A},\mathcal{B}).
\end{equation*}
The neutrino oscillation probability calculated from the former is a periodic function dependent on sidereal time, while the oscillation probability calculated from the latter equals the average value of this periodic function,
\begin{equation}
    P(\mathcal{A},\mathcal{B})\simeq\frac{1}{T}\int_0^T P_{\bar{\nu}_e\rightarrow\bar{\nu}_e} (a^\mu_L,c^{\mu\nu}_L,t)dt.
\end{equation}
The neutrino beam traverses all azimuths in $T=2\pi/\omega=\text{23 h 56 min}$ due to the Earth's rotation.

\bibliography{main}

\end{document}